\documentclass[journal]{IEEEtran}
\usepackage{amsmath}
\usepackage{subfigure} 
\usepackage{lineno,hyperref}
\usepackage{xcolor}
\usepackage{extarrows}
\usepackage{algorithm}
\usepackage{algorithmic}
\usepackage{amssymb}
\usepackage{bm}
\usepackage{graphicx}
\usepackage[justification=centering]{caption}

\begin{document}
\title{Sparse Bayesian Learning Using Approximate Message Passing with Unitary Transformation}

\author{ Man~Luo, Qinghua~Guo\textsuperscript{1} \\
School of Electrical, Computer and Telecommunications Engineering, University of Wollongong\\
qguo@uow.edu.au\\
\thanks{\textsuperscript{1} Corresponding: Q. Guo. This is part of our work in progress (presented in IEEE APWCS 2019).}}
\markboth{}%
{Shell \MakeLowercase{\textit{et al.}}: Bare Demo of IEEEtran.cls for IEEE Journals}

\maketitle

\begin{abstract}
Sparse Bayesian learning (SBL) can be implemented with low complexity based on the approximate message passing (AMP) algorithm. However, it is vulnerable to `difficult' measurement matrices as AMP can easily diverge. Damped AMP has been used to alleviate the problem at the cost of slowing the convergence speed. In this work, we propose an SBL algorithm based on the AMP with unitary transformation (UTAMP), where the shape parameter of the hyperprior is tuned automatically. It is shown that, compared to the state-of-the-art AMP based SBL algorithm, the proposed UTAMP-SBL is much more robust and much faster, leading to remarkably better performance. It is shown that in many cases, UTAMP-SBL can approach the support-oracle bound closely.  
\end{abstract}


\IEEEpeerreviewmaketitle

\section{Introduction}
In this work, we consider a sparse Bayesian learning (SBL) problem with the following model
\begin{equation}
 \mathbf{y} = \mathbf{Ax} + \mathbf{w}\\
\label{y=ax+W}
\end{equation}
where $\mathbf{y}$ is an observation vector with length $M$, $\mathbf{A}$ is a known measurement matrix with a size of $M \times N$, $\mathbf{x}$ is a length-$N$ sparse vector to be recovered, and $\mathbf{w}$ denotes a Gaussian noise vector with mean zero   and  covariance  matrix $\lambda^{-1}\mathbf{I}$. It is assumed that the elements of   $\mathbf{x}$ are independent and identically distributed, i.e., $p(\mathbf{x})=\prod_{n}p(x_n)$. 
The sparsity promoting prior  $p(x_n)=\int \mathcal{N} (  {x_n}| 0 , \gamma_n^{-1}) p(\gamma_n)d\gamma_n$, where $\mathcal{N} (  {x_n}| 0 , \gamma_n^{-1}) $ denotes a Gaussian density with mean zero and variance $\gamma_n^{-1}$, and
$p(\gamma_n)$ is a proper hyperprior. 

The approximate message passing (AMP) algorithm was developed for compressive  sensing based on the loopy belief propagation \cite{donoho2009message}.  
AMP has low complexity and its performance can be rigorously characterized by a scalar state evolution in the case of large i.i.d (sub)Gaussian matrix $\mathbf{ A }$ \cite{javanmard2013state}. 
 However, for a generic $\mathbf{A}$, the convergence of AMP cannot be guaranteed, e.g., AMP can easily diverge for non-zero mean, rank-deficient, correlated, or ill-conditioned matrix $\mathbf{A}$ \cite{rangana2019convergence}.
To address this problem, many AMP variants have been proposed, such as the damped AMP \cite{rangana2019convergence}, the swept AMP \cite{manoel2014sparse}, and GAMP with an adaptive damping and mean-removal procedure \cite{vila2015adaptive}. More effective variants include the AMP with unitary transform (UTAMP) \cite{guo2015approximate} proposed in 2015, and the vector AMP \cite{rangan2017vector} and the orthogonal AMP \cite{ma2017orthogonal} proposed in 2016. In particular, UTAMP was derived based on a unitary transform of model (\ref{y=ax+W}), and it converges for any matrix $\mathbf{ A}$ in the case of Gaussian priors \cite{guo2015approximate}.   

AMP and its variants have been used for low complexity implementation of SBL \cite{al2014sparse}. By using AMP to implement the E-step in the expectation maximization (EM) based SBL, a significant reduction in complexity can be achieved. 
However, AMP-SBL can diverge easily for a difficult matrix $\mathbf{ A }$, and exhibits poor performance. In \cite{al2018gamp},  a GAMP based SBL algorithm (GGAMP-SBL) was proposed where the convergence is improved through damping but at the expense of slowing the convergence speed.

In this work, we derive a new SBL algorithm based on UTAMP (called UTAMP-SBL) to achieve low complexity robust SBL by taking advantage of the robustness and low complexity of UTAMP. In UTAMP-SBL, a Gamma distribution is chosen as the hyperprior for precision $\gamma_n$, and the shape parameter of the Gamma distribution is automatically tuned during the iteration. 
It is shown that, compared to the state-of-the-art AMP based SBL algorithm GGAMP-SBL \cite{al2018gamp}, UTAMP-SBL can deliver remarkably better performance in terms of robustness, speed and recovery performance. It is observed that, in many cases with difficult measurement matrices, UTAMP-SBL can still approach the support-oracle bound closely. 

Notations: 
Boldface lowercase letters and uppercase symbols represent column vectors and matrices, respectively. Scalars are represented by non-boldface letters. 
$\mathbf{1}$ and $\mathbf{0}$ represent an all-one column vector and an all-zero column vector with proper sizes.
The $n$-th element of vector $\mathbf{c}$ is denoted by $c_n$. 
Diag($\mathbf{a}$) returns a diagonal matrix with the elements of $\mathbf{a}$ on its diagonal. 
$(\mathbf{A})_D$ returns a diagonal matrix by forcing the off-diagonal elements of $\mathbf{A}$ to zero. $\mathcal{N} ( \mathbf{x}| \bm{\mu},\bm{\Sigma} )$ denotes a Gaussian density of  $\mathbf{x}$ with mean $\bm{\mu}$ and covariance   $\bm{\Sigma}$, and ${\mathrm{Ga}(\bm{\gamma}| \epsilon,\eta)}$ denotes a Gamma distribution with shape parameter $\epsilon$ and rate parameter $\eta$. Let $(\cdot)^H$, $ \left\|\cdot \right\|$, $ \left|\cdot \right|$ and $\delta \left(\cdot \right)$ denote the (conjugate) transpose, the $l_2$ norm, the element-wise magnitude squared operations and the Dirac delta function, respectively.  The notation $\left\langle f(\mathbf{x}) \right\rangle _{q(\mathbf{x})}$ denotes the expectation of the function $f(\mathbf{x})$ with respect to probability density $q(\mathbf{x})$.

\section{ AMP with Unitary Transformation } 
 \begin{algorithm}
 	\caption{Vector Stepsize AMP}
 	Initialize $\bm{\tau}_x^{(0)}>0$ (with elements larger than 0) and $\mathbf{x}^{(0)}$. Set $\mathbf{s}^{(-1)}=\mathbf{ 0 }$ and $ t=0$.\\
 	\textbf{Repeat}   
 	\begin{algorithmic}[1]
 		\STATE $\bm{\tau}_p$ = $ | \mathbf{A} |^2  \bm{\tau}^t_x$\\
 		\STATE $\mathbf{p}= \mathbf{ \mathbf{A} x^t} - \bm{\tau}_{p}  \cdot  \mathbf{s}^{t-1} $\\
 		\STATE $\bm{\tau}_s = \mathbf{1} ./ (\bm{\tau}_p+\lambda^{-1} \mathbf{1}) $\\
 		\STATE $ \mathbf{s}^t=\bm{\tau}_s \cdot  (\mathbf{y}-\mathbf{p}) $\\
 		
 		\STATE 	$ \mathbf{1} ./\bm{\tau}_q =  | \mathbf{A}^H |^2   \bm{\tau}_s  $\\
 		\STATE $ \mathbf{q} = \mathbf{x}^t + \bm{\tau}_q \cdot  \mathbf{A}^H   \mathbf{s}^t $\\
 		\STATE 	$\bm{\tau}_x^{t+1}$ = $ \bm{\tau}_q  \cdot   g_{x}' ( \mathbf{q}, \bm{\tau}_q)  $\\
 		\STATE 	$\mathbf{x}^{t+1} = g_{x}  ( \mathbf{q}, \bm{\tau}_q)$	
 		\STATE $t=t+1$
 	\end{algorithmic}
 	\textbf{Until terminated}   
 	\label{the vector stepsize AMP-table}
 \end{algorithm}
The UTAMP algorithm, inspired by \cite{guo2013iterative}, was derived based on the vector stepsize AMP algorithm shown in Algorithm~\ref{the vector stepsize AMP-table} and a unitary transform of model (\ref{y=ax+W}) \cite{guo2015approximate}. In AMP and UTAMP, the function $g_x(\mathbf{q}, \bm{\tau}_q )$ returns a column vector whose $n$-th
element, denoted as $[ g_x(\mathbf{q}, \bm{\tau}_q ) ]_n$, is given by 
\begin{equation}
 [g_x(\mathbf{q}, \bm{\tau}_q ) ]_n
 =
 \frac{\int x_n p(x_n) \mathcal{N} (x_n | q_n, \tau_{q_n})  d x_n }{\int  p(x_n) \mathcal{N} (x_n | q_n, \tau_{q_n})  d x_n }.
 \label{g_x}
\end{equation}
Equation (\ref{g_x}) can be interpreted as the minimum mean square error (MMSE) estimation of $x_n$ based on the following model
\begin{equation}
q_n 
=
x_n + \varpi 
\label{q_n}
\end{equation}
where    $\varpi $ is a Gaussian noise with mean zero and variance $\tau_{q_n}$. 
The function $g_x'(\mathbf{q},\bm{\tau}_q)$ returns a column vector and the $n$-th element is denoted by $[ g_x(\mathbf{q}, \bm{\tau}_q ) ]_n$, where the derivative is with respect to $q_n$. It is not hard to show that  $ {\tau}_{q_n} [ g_x(\mathbf{q}, \bm{\tau}_q ) ]_n$  is the a posterior variance of $x_n$ with model (\ref{q_n}). Note that $g_x(\mathbf{q},\bm{\tau}_q)$ can also be changed for MAP (maximum a posterior) estimation of $\mathbf{x}$.
 
The derivation of UTAMP is briefly introduced in the following. As any matrix $\mathbf{ A }$ can have its singular value decomposition (SVD)  $\mathbf{A}= \mathbf{U \Lambda V}$, a unitary transformation with $\mathbf{U}^H$ to (\ref{y=ax+W}) can be performed, yielding
\begin{equation}
\mathbf{r=\Lambda V x}+\bm{\omega}
\label{r=uy}
\end{equation}
where $\mathbf{r=U^H y}$, and $\bm{\omega} = \mathbf{U^H w }$  is still a zero-mean  Gaussian noise vector with the  same covariance matrix  $\lambda^{-1} \mathbf{I}$.
 
It is not hard to verify that 
 \begin{equation}
  |\mathbf{ C }|^2 \mathbf{ d } = (\mathbf{ C } Diag(\mathbf{ d })  \mathbf{ C }^H )_D \mathbf{1}.
  \label{math}
 \end{equation}
Now suppose we have a variance vector $\bm{\tau}^t_x$. According to Line 1 in the vector stepsize AMP and using (\ref*{math}), we have 
\begin{equation}
\bm{\tau}_p = (\bm{\Lambda} \mathbf{V} { Diag}(\bm{\tau}^t_x) \mathbf{V}^H \bm{\Lambda}^H )_D \mathbf{1}.
\label{tau_p}
\end{equation}
   \begin{algorithm}
	\caption{UTAMP}
	Unitary transform: $\mathbf{r=U}^H \mathbf{ y }=\mathbf{\Lambda V x} +\bm{\omega}$, where $\mathbf{A=U \Lambda V}$.\\
	Define vector $\bm{\lambda_p}=\mathbf{ \Lambda \Lambda}^H \textbf{1}$.\\  
    Initialize ${\tau}_x^{(0)}>0$  and $\mathbf{x}^{(0)}$. Set $\mathbf{s}^{(-1)}=\mathbf{ 0 }$ and $t=0$\\
    \textbf{Repeat}   
 	\begin{algorithmic}[1]

	    \STATE $\quad		\bm{\tau}_p$ = $ \tau^t_x  \bm{\lambda}_p$\\
		\STATE $\quad		\mathbf{p}= \mathbf{\Lambda V x}^t - \bm{\tau}_{p} \cdot  \mathbf{s}^{t-1} $\\
		\STATE $\quad		\bm{\tau}_s = \mathbf{1}./ (\bm{\tau}_p+\lambda^{-1} \mathbf{1}) $\\
		\STATE $\quad		\mathbf{s}^t= \bm{\tau}_s \cdot (\mathbf{r}-\mathbf{p}) $\\
		
		\STATE 	$\quad		1/\tau_q = ({1}/{N}) \bm{   \lambda_p}^H \bm{\tau }_s   $\\
		\STATE $\quad		 \mathbf{q} = \mathbf{x}^t + \tau_q (\mathbf{V}^H \mathbf{\Lambda}^H \mathbf{s}^t)$\\
		\STATE 	$\quad		\tau_x^{t+1}$ = $ (\tau_q  /N)  \mathbf{1}^H   g_{x}' ( \mathbf{q}, \tau_q) $\\
		\STATE 	$\quad		\mathbf{x}^{t+1} = g_{x}  ( \mathbf{q}, \tau_q)$	
 		\STATE 	$\quad	      t=t+1$
	\end{algorithmic}
\textbf{Until terminated}   
\label{UTAMP-table}
\end{algorithm}
We can find that if $Diag(\bm{\tau}_x^t)$ is a scaled identity matrix, the computation of  ($\ref{tau_p}$) can be significantly simplified. This motives the replacement of $\bm{\tau}_x^t$ with ${\tau}_x^t \mathbf{1}$ where ${\tau}_x^t$ is the average of the elements of $\bm{{\tau}}_x^t$. So ($\ref{tau_p}$) is reduced to 
 \begin{equation}
 \bm{\tau}_p= \tau_x^t \bm{\Lambda \Lambda}^H \mathbf{1}
 \end{equation}
which is Line 1 of the UTAMP algorithm. Lines 2, 3 and 4 of UTAMP can be obtained according to Lines 2, 3 and 4 of the vector stepsize AMP by simply replacing $\mathbf{A}$ with $ \bm{\Lambda} \mathbf{ V }$. According to (\ref{math}) again, 
Line 5 of the vector stepsize AMP with matrix $\bm{\Lambda} \mathbf{V}$ can be represented as
\begin{equation}
\mathbf{1}./\bm{\tau}_q =( \mathbf{V}^H \bm{\Lambda}^H {\rm Diag} (\bm{\tau}_s)\bm{\Lambda} \mathbf{V} )_D \mathbf{1}.
\label{tau_q}
\end{equation}
We then replace the diagonal matrix $\bm{\Lambda}^H {\rm Diag} (\bm{\tau}_s) \bm{\Lambda} $ with a scaled identity matrix  $\beta \mathbf{I} $ where $\beta$ is the average of the diagonal elements of $\bm{\Lambda}^H {\rm Diag} (\bm{\tau}_s) \bm{\Lambda} $, i.e.,
 \begin{equation}
 \beta
 =
 (1/N) \mathbf{1}^H \bm{\Lambda} \bm{\Lambda}^H \bm{\tau}_s.
 \end{equation}  
 Hence (\ref{tau_q}) is reduced to Line 5 of the UTAMP algorithm. Line 6 can be obtained from Line 6 of the vector stepsize AMP by replacing $\mathbf{A=U \Lambda V}$ with $\bm{\Lambda} \mathbf{V}$. Compared with Line 7 in the vector stepsize AMP, an additional average operation is performed in Line 7 in UTAMP to meet the requirement of a scalar $\tau_x^t$ in Line 1. We note that the average operation is not necessarily in Line 7 as we can also put the additional average operation in Line 1. Line 8 in UTAMP is the same as Line 8 of the vector stepsize AMP except that $\tau_q$ is a scalar. The UTAMP algorithm is summarized in Algorithm \ref{UTAMP-table}.  
 
\textbf{Remarks:} It is worth pointing out that UTAMP is not equivalent to the vector step size AMP due to the approximations made in the derivation. Interestingly, it is these approximations that make UTAMP much more robust than AMP. 

\section{Sparse Bayesian leaning with UTAMP}
  
\begin{figure}
	\centering
	\includegraphics[width=1\linewidth]{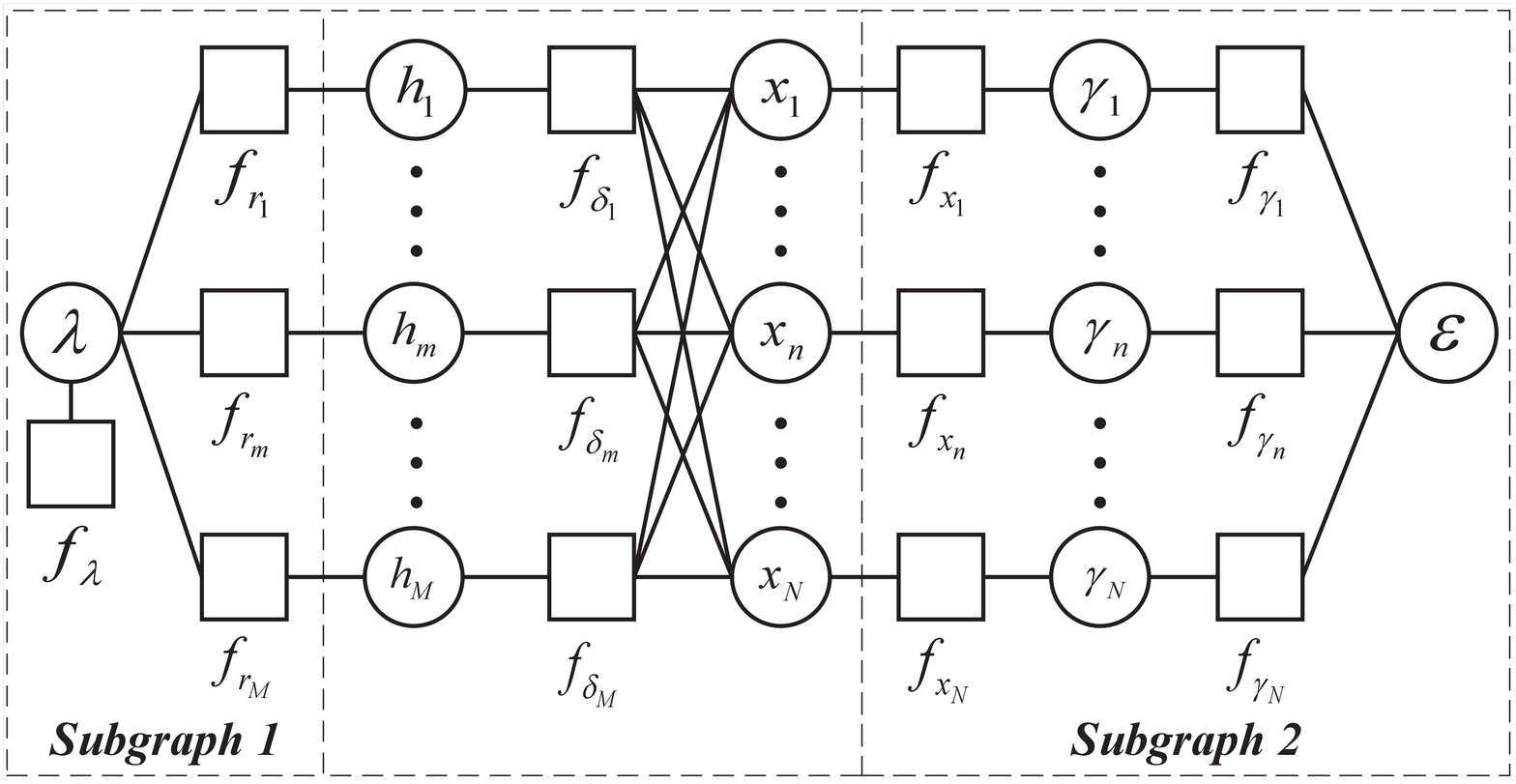}
	\centering
	\caption{Factor graph for deriving UTAMP-SBL.}
	\label{fig:factor graph}
\end{figure}

In this Section, the SBL algorithm UTAMP-SBL is proposed based on the UTAMP algorithm, where we assume that the noise precision $\lambda$ is unknown. The UTAMP-SBL algorithm is derived using the factor graph representation based on model (\ref*{r=uy}). The factor graph is shown in Fig.~\ref{fig:factor graph}, where $f_{r_m}({r_m}| h_m, \lambda ) =\mathcal{N} (r_m | h_m, \lambda^{-1})$, $ f_{\delta_m}(h_m|{\bm{\mathbf{x}} })  =  \delta(h_m-(\bm{\Lambda V})_m \mathbf{x} ) $ with $(\bm{\Lambda V})_m$ being the $m$-th row of matrix $\mathbf{ \bm{\Lambda V} } $,
$f_{\lambda}(\lambda)$ denotes the prior of the noise precision  $\lambda$,
$f_{x_n}({x_n}| \gamma_n)=\mathcal{N} (x_n |  0, \gamma_n^{-1})$ denotes the Gaussian prior of $x_n$, and $f_{\gamma_n}(\gamma_n) =\mathrm{Ga}(\lambda|\epsilon,\eta) $ is the hyperprior. UTAMP-SBL is derived by incorporating the UTAMP algorithm to the message passing in the factor graph shown in Fig.~\ref{fig:factor graph}, where the whole graph is divided into 3 subgraphs. The subgraph in the middle is mainly handled by UTAMP, and the message computations in Subgraphs 1 and 2 are detailed in the following.

\subsection{Message Computations in Subgraph 1}

\subsubsection{Backward Message Passing}
According to the derivation of AMP using loopy belief propagation, UTAMP provides the message $m_{{h_m}\rightarrow f_{r_m}} (h_m) = \mathcal{N} (h_m |  p_m, \tau_{p_m})$ where the mean $p_m$ and the variance $\tau_{p_m}$ are given in Lines 1 and 2 of the UTAMP algorithm (Algorithm 2), which are also Lines 1 and 2 in the UTAMP-SBL algorithm (Algorithm 3). The mean field (MF)  rule  is used at the function nodes $\left\lbrace  f_{r_m} \right\rbrace $. Hence, we need to compute the belief $b(h_m )$, i.e., 
$b(h_m) \propto  m_{{h_m}\rightarrow f_{r_m}} (h_m) m_{f_{r_m}\rightarrow h_m} (h_m)$.  
Later we will see that $m_{f_{r_m}\rightarrow h_m} (h_m) \propto\mathcal{N} (h_m | r_m, \hat{ \lambda}^{-1})$ where $\hat{ \lambda}^{-1}$ is an estimate of $\lambda$ (in the last iteration), and its computation is delayed to (\ref{n1}). Hence  $b(h_m )$ is Gaussian  i.e., $b(h_m)  = \mathcal{N} (h_m | \hat{h}_m, v_{h_m})$,
where 
\begin{equation}
	v_{h_m} = ({ {1}/{\tau_{p_m}} +\hat{ \lambda} })^{-1} \\
 ,~~ 
 \hat{h}_m =  v_{h_m} ( {r_m} \hat{ \lambda} +{p_m}/{\tau_{p_m}})
\end{equation}
leading to Lines 3 and  4 of the UTAMP-SBL algorithm. The message from $f_{r_m}$ to ${ \lambda}$ is calculated by the MF rule, which reads
\begin{equation}
\begin{aligned}
m_{f_{r_m}\rightarrow \lambda} (\lambda) &= \exp \left\{  \left\langle {\log  f_{r_m} ( {{r_m| h_m, \lambda^{-1}}}  )} \right\rangle_{b\left( {h_m} \right) } \right\}\\
&  \propto \sqrt{\lambda} \exp \left\{   { -\frac{\lambda}{2} } { ( || r_m - \hat{h}_m  ||^2 + v_{h_m}  ) } \right\} .
\end{aligned}
\end{equation}

\subsubsection{Forward Message Passing}
According to the MF rule, the message $m_{f_{r_m}\rightarrow h_m} (h_m) $ reads
\begin{equation}
\begin{aligned}
m_{f_{r_m}\rightarrow h_m} (h_m)
& =\exp \left\{  \left\langle {\log  f_{r_m} ( {{r_m| h_m, \lambda^{-1}}}  )} \right\rangle_{b\left( {\lambda} \right) } \right\}\\
&\propto \mathcal{N} (h_m |  r_m, \hat{ \lambda}^{-1})
\label{M_frm_to_hm}
\end{aligned}
\end{equation}
where $\hat{ \lambda}= \left\langle \lambda \right\rangle _{b(\lambda)}$. 
With the prior density $f_{\lambda} (\lambda)  \propto \frac{1}{\lambda}$,
\begin{equation}
\begin{aligned}
b(\lambda)& \propto f_{\lambda} (\lambda)  \prod_{m} m_{f_{r_m}\rightarrow \lambda} (\lambda) \\
&  \propto  {\lambda}^{\frac{M}{2}-1} \exp \left\{   { -\frac{\lambda}{2} } \sum_{m} { \left( || r_m - \hat{h}_m  ||^2 + v_{h_m}  \right) } \right\}. 
\end{aligned}
\end{equation}
Hence  
\begin{equation}
\hat{ \lambda}  =  {M}/{\sum_{m} { \left( || r_m - \hat{h}_m  ||^2 + v_{h_m}  \right) }},\label{n1}
\end{equation} 
which is Line 5 of the UTAMP-SBL algorithm.

   \begin{algorithm}
	\caption{  UTAMP-SBL}
	Unitary transform :$\mathbf{r=U}^H \mathbf{ y }=\mathbf{\Lambda V x} +\bm{\omega}$, where $\mathbf{A=U \Lambda V}$.\\
	Define vector  $\bm{\lambda_p}= \mathbf{\Lambda \Lambda}^H \mathbf{1}$.\\  
	Initialization: ${\tau}_x^{(0)}=1$, $\hat{\mathbf{x}}^{(0)}=\textbf{0}$, $\hat{\epsilon}=0.001$, $\hat{{\bm{\gamma}}}^{(0)}=\textbf{1}$, $\hat{ \lambda}^{(-1)}=1$, $\mathbf{s}^{(-1)}=\mathbf{ 0 }$, and $t=0$.
	
	\textbf{Do}     
	\begin{algorithmic}[1]
		\STATE $ 		\bm{\tau}_p$ = $ \tau^t_x  \bm{\lambda}_p$\\
		\STATE $ 		\mathbf{p}= \mathbf{\Lambda V} \hat{\mathbf{x}}^t - \bm{\tau}_{p} \cdot  \mathbf{s}^{t-1} $\\
		
		\STATE  $        \mathbf{v}_h=\mathbf{1}./(\mathbf{1}./\bm{\tau}_p +\hat{ \lambda}^{t-1}\mathbf{1})$\\      
		\STATE $         \mathbf{\hat{h}}=\mathbf{v}_h \cdot (\hat{ \lambda}^{t-1} \mathbf{r} + \mathbf{p}./ \bm{\tau}_p )$\\
		\STATE  $\hat{ \lambda}^t =   {M}/   ( {  ||  \mathbf{r}-  \mathbf{\hat{h}}  ||^2   + \mathbf{1}^H\mathbf{v}_h  }  ) $;\\	
		\STATE $     \bm{\tau}_s = \mathbf{1}./ (\bm{\tau}_p+  ({\hat{ \lambda}^t})^{-1}\mathbf{1}) $\\
		\STATE $      \mathbf{s}^t= \bm{\tau}_s \cdot (\mathbf{r}-\mathbf{p}) $\\
		
		\STATE 	$    1/\tau_q = ({1}/{N}) \bm{   \lambda_p}^H \bm{\tau_s   }$\\
		\STATE $     \mathbf{q} =\hat{\mathbf{x}} ^t + \tau_q (\mathbf{V}^H \mathbf{\Lambda}^H \mathbf{s}^t)$\\

		\STATE 	$   \tau_x^{t+1}=  (\tau_q/N) \mathbf{1}^H(\mathbf{1}. /(\mathbf{1}+\tau_q  \hat{{\bm{\gamma}}}^t ))$
		\STATE 	$   \hat{\mathbf{x}}^{t+1} = 	 \mathbf{q}   ./(\mathbf{1}+\tau_q \hat{{\bm{\gamma}}}^t )$\\
		\STATE $   \hat{{\gamma}}_n^{t+1}  = ( {2\hat{\epsilon} +1})/( |{{\hat{x}}_n^{t+1} }|^2 +\tau_x^{t+1}), n=1, ...,N.$ \\
		(Note: $ \hat{{\gamma}}_n^{t+1}  = ( {\hat{\epsilon} +1})/( |{{\hat{x}}_n^{t+1} }|^2 +\tau_x^{t+1})$ for complex case)\\
		\STATE $\hat{\epsilon}=\frac{1}{2}\sqrt{\log(\frac{1}{N}\sum_{n}{\hat{\gamma}_n^{t+1}})-\frac{1}{N}\sum_{n}{\log{\hat{\gamma}}_n^{t+1}}}$  
		\STATE   $ t=t+1 $  
	\end{algorithmic}
	\textbf{while}	({$   \left\| \hat{\mathbf{x}}^{t+1} - \hat{\mathbf{x}}^{t} \right\|^2/ \left\|  \hat{\mathbf{x}}^{t+1}  \right\|^2  > \delta_x$} and $t<t_{max}$)
	\label{UTAMP_SBL-optimized}
\end{algorithm}

\subsection{Message Computations in Subgraph 2}
\subsubsection{Forward Message Passing}
According to the derivation of AMP and UTAMP, UTAMP produces the message   $m_{{x_n}\rightarrow f_{x_n}} (x_n) \propto \mathcal{N} (x_n |  q_n, \tau_q)$ with mean $q_n$ and variance $\tau_q$, which correspond to Lines 5 and 6 of the UTAMP algorithm and Line 8 and Line 9 of the UTAMP-SBL algorithm. The MF rule is also used at nodes $\left\lbrace  f_{x_n}\right\rbrace $. Hence, we need to compute the belief $b(x_n )$, i.e., $b(x_n) \propto m_{{x_n}\rightarrow f_{x_n}} (x_n)  m_{f_{x_n}\rightarrow x_n} (x_n)$.
The message $m_{f_{x_n}\rightarrow x_n} (x_n) \propto  \mathcal{N}  ( {{x_n| 0, \hat{ \gamma}_n^{-1}}}  )$  will be defined in (\ref{M_fx_x}),  where $\hat{ \gamma}_n= \left\langle \gamma_n \right\rangle _{b(\gamma_n)}$. So $b(x_n)=\mathcal{N} (  {x_n}  | \hat{x}_n, \tau_x)$ with
\begin{equation}
\tau_{x_n}  = \left( {1}/{\tau_q} +\hat{ \gamma}_n \right) ^{-1}, ~~ \hat{ x }_n  =  {q_n} /{(1+ \tau_q \hat{ \gamma}_n )} 
\end{equation}
leading to  Lines 10 and 11 of the UTAMP-SBL algorithm (noting that $
\tau_x$ is the average of $\{\tau_{x_n}\}$).

By the MF rule, the message $m_{f_{x_n}\rightarrow {\gamma_n}} (\gamma_n) $ from $f_{x_n }$ to $\gamma_n$ is calculated by
\begin{eqnarray}
\begin{aligned}
m_{f_{x_n}\rightarrow {\gamma_n}} (\gamma_n) 
&= \exp \left\{  \left\langle  { {\log  f_x ( {{x_n| 0, \gamma_n^{-1}}}  )}   } \right\rangle _{b\left( {x_n} \right)   } \right\}  \\
&\propto \sqrt \gamma_n  \exp \left\{  -0.5{\gamma_n} {  ( {  |\hat{x}_n|^2 + \tau_x} )   } \right\}
\label{M_fx_ga}
\end{aligned}
\end{eqnarray}
As defined later in (\ref{M_fgn_to_gn})   $m_{f_{\gamma_n}\rightarrow {\gamma_n}} (\gamma_n) \propto {\gamma_n}^{\hat{\epsilon}-1}  \exp \left\{  -\frac{\gamma_n}{2} {  ( {|\hat{x}_n|^2 + \tau_x} )   } \right\}$  where $\hat{ \epsilon}=\left\langle \epsilon \right\rangle _{b(\epsilon)}$,  the belief $b(\gamma_n )$ is scaled to a Gamma distribution  i.e.,
\begin{equation}
\begin{aligned}
b(\gamma_n)
&\propto   m_{f_{\gamma_n}\rightarrow \gamma_n} (\gamma_n)  m_{f_{x_n}\rightarrow \gamma_n} (\gamma_n)  \\
&\propto    {\gamma_n}^{\hat{\epsilon} - \frac{1}{2}}  \exp \left\{  -\frac{\gamma_n}{2} {  ( {|\hat{x}_n|^2 + \tau_x  +2\eta} )   } \right\}.
\end{aligned}
\end{equation}
Hence $\hat{ \gamma}_n = \frac{2\hat{\epsilon}+1}{ 2\eta +{{|\hat{x}_n|^2 + \tau_x}  }}$. 
Here we set $\eta=0$, and $\hat{ \gamma}_n$ is reduced to $  \frac{2\hat{\epsilon}+1}{{  {|\hat{x}_n|^2 + \tau_x}  }}$ 
, which leads to Line 12 of the UTMP-SBL algorithm. 


According to the MF rule, $m_{f_{\gamma_n} \rightarrow {\epsilon}}(\epsilon)=\frac{\eta^\epsilon}{\Gamma(\epsilon)} \exp \left\{  (\epsilon-1) \left\langle \log{   { \gamma_n } } \right\rangle_{b ( {\gamma_n} ) } - \eta \hat{ \gamma}_n \right\} $. Thus, 
\begin{equation}
\begin{aligned}
b(\epsilon)
&\propto \prod_{n} m_{f_{\gamma_n} \rightarrow {\epsilon}} (\epsilon) \\
&\propto
 \frac{\eta^{N \epsilon }}{ \left( \Gamma(\epsilon)\right) ^N  } \exp \left\lbrace  (\epsilon-1) \sum_{n}    \left( \left\langle \log{   { \gamma_n } } \right\rangle_{b ( {\gamma_n} ) } - \eta \hat{ \gamma}_n \right) \right\} 
\end{aligned}
\label{b(epsilon)}
\end{equation}

\subsubsection{Backward Message Passing}
According to the MF rule, the message $m_{f_{\gamma_n} \rightarrow {\gamma_n}} (\gamma_n)$ is calculated as
\begin{equation}
\begin{aligned}
m_{f_{\gamma_n}\rightarrow \gamma_n} (\gamma_n)
 =\exp \left\{  \left\langle {\log  f_{\gamma} ( {{\gamma_n| \epsilon,  \eta}}  )} \right\rangle_{b\left( {\epsilon} \right) } \right\}
\propto \mathcal{N} (\gamma_n |  \hat{ \epsilon}, \eta)
\label{M_fgn_to_gn}
\end{aligned}
\end{equation}
where $\hat{ \epsilon}=\left\langle \epsilon \right\rangle _{b(\epsilon)}$.  
However, it is difficult to calculate $\hat{\epsilon}$. To circumvent this problem, we may use $\epsilon'=\mathrm{argmax} _{\epsilon} b(\epsilon)$ to replace $\hat{\epsilon}$, and $\epsilon'$  can be obtained iteratively with the following equation \cite{fink1997compendium} 

\begin{figure}
	\centering
	\subfigure[$\rho$=0.1]{
		\includegraphics[width=1\linewidth]{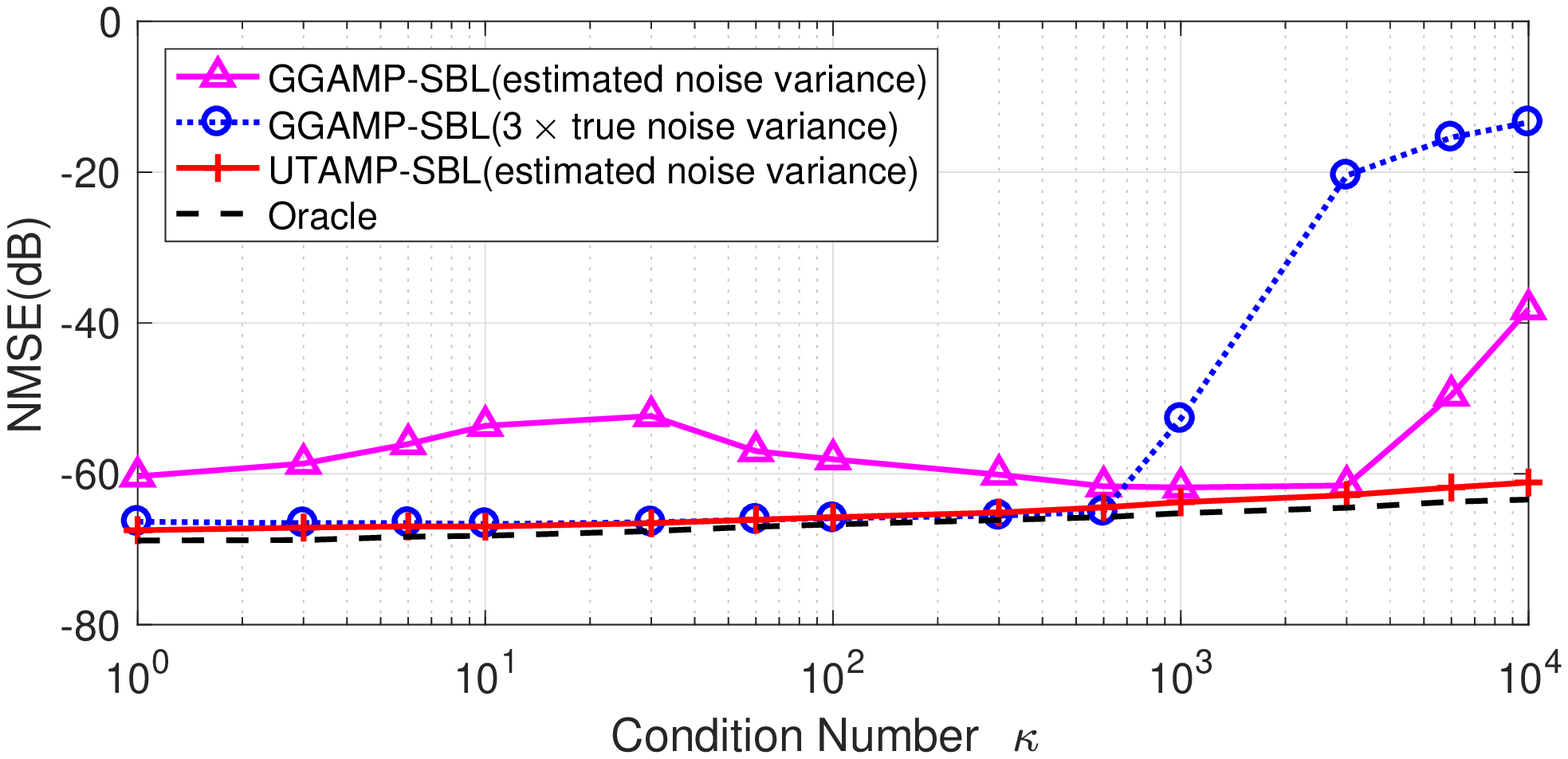}}
	\centering
	\subfigure[$\rho$=0.3]{
		\includegraphics[width=1\linewidth]{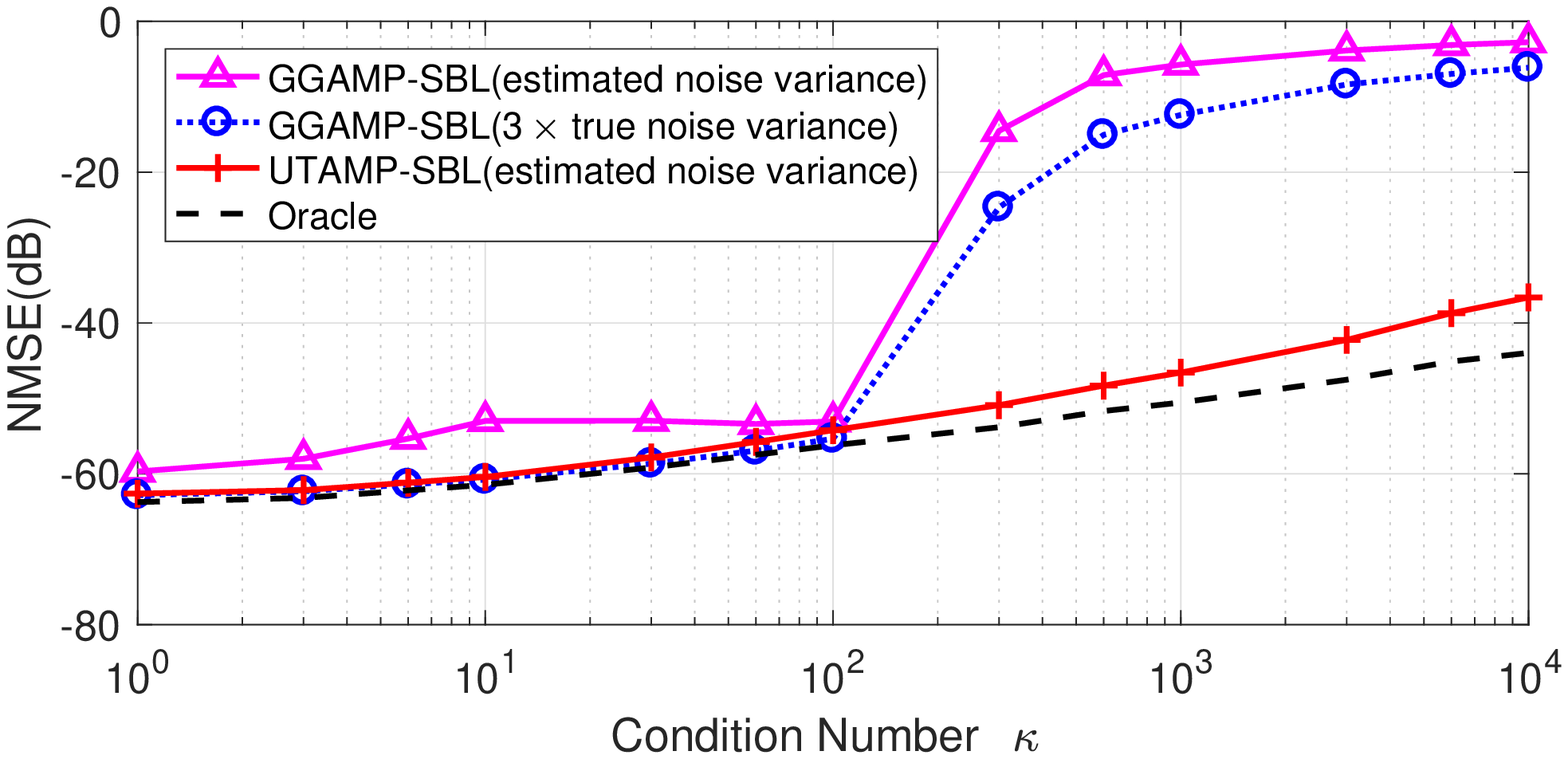}}
	\caption{Performance under ill-conditioned matrices.}
	\label{conditionresult}
\end{figure}

\begin{figure}
	\centering
	\subfigure[$\rho$=0.1]{
		\includegraphics[width=1\linewidth]{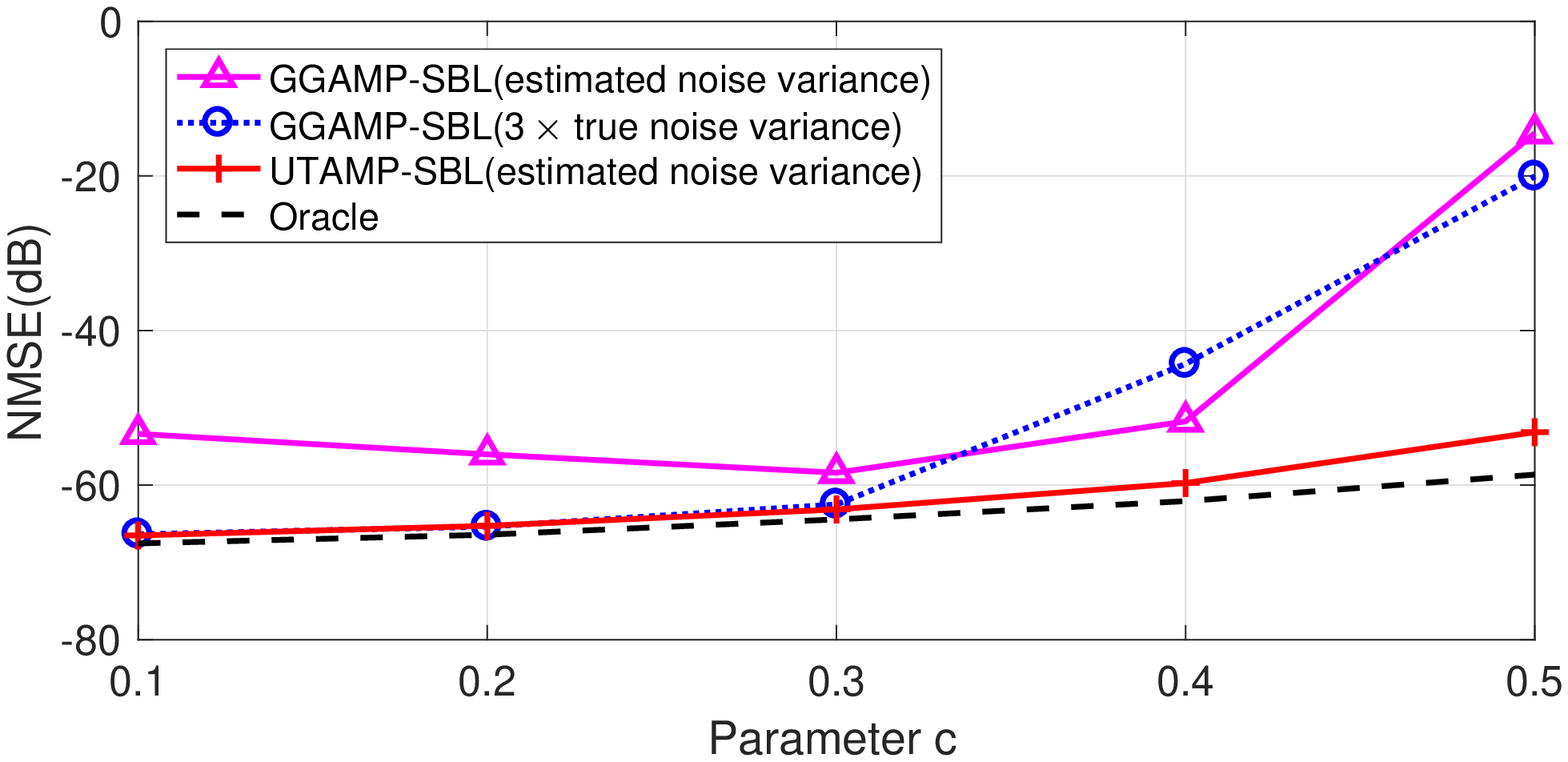}}
	\centering
	\subfigure[$\rho$=0.3]{
		\includegraphics[width=1\linewidth]{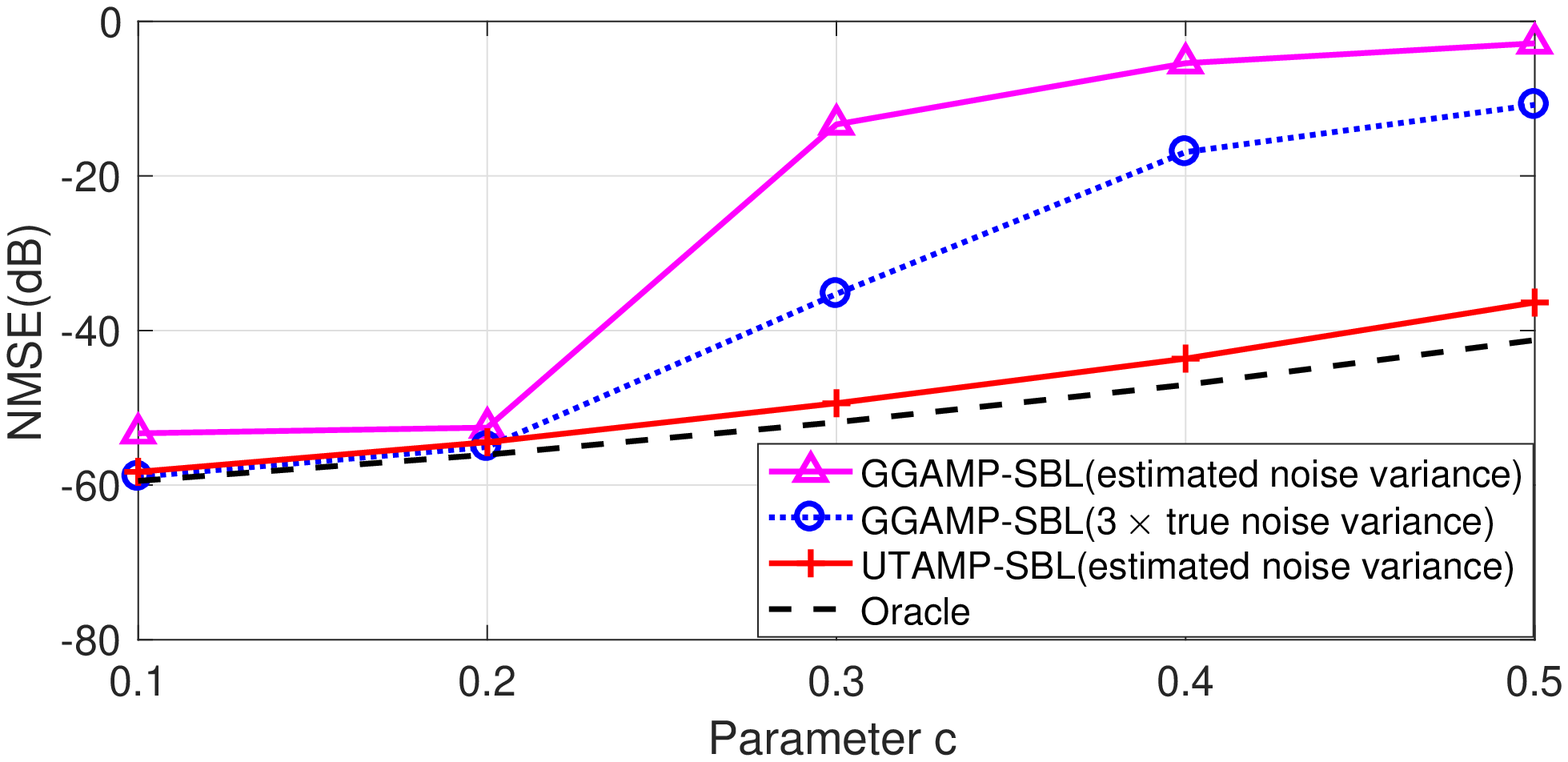}}
	\caption{Performance under correlated matrices.}
	\label{correlationresult}
\end{figure}

 
\begin{equation}
\begin{aligned}
\frac{1}{ {\epsilon}_{new}} 
&= \frac{1}{ {\epsilon}} + 
  \frac{  g (\epsilon)}{{ {\epsilon}}^2 (1/  {\epsilon} -\Psi'( {\epsilon}) )  }
\end{aligned}
\end{equation}
where  
\begin{equation}
\begin{aligned}
  g (\epsilon)&= \frac{1}{N}{\sum_{n} \log  \hat{ \gamma}_n} - \Psi( {\epsilon}) +\log  \left( \frac{\epsilon N }{\sum_{n}  \hat{ \gamma}_n} \right) \\ 
 & +  \Psi  ( \hat{ \epsilon} +0.5) -\log  ( \hat{ \epsilon} +0.5 ),
   \end{aligned}
   \nonumber
\end{equation}
$\Psi$ is the digamma function and $\Psi'$ is the derivative of the digamma function. We find the following simple but more effective equation to update $\hat{\epsilon}$, where iteration is not required and  $\hat{\epsilon}$ can be calculated straightaway
\begin{equation}
\hat{\epsilon}=
\frac{1}{2}\sqrt{\log(\frac{1}{N}\sum_{n}{\hat{\gamma}_n})-\frac{1}{N}\sum_{n}{\log{\hat{\gamma}}_n}},
\label{eq_eps_update}
\end{equation} 
which is Line 13 of the UTAMP-SBL algorithm. 
By the MF rule, the message $m_{f_{x_n}\rightarrow {x_n}} (x_n) $ is updated by
\begin{equation}
\begin{aligned}
m_{f_{x_n}\rightarrow x_n} (x_n) 
&= \exp \left\{  \left\langle  { {\log  f_x ( {{x_n| 0, \gamma_n^{-1}}}  )}   } \right\rangle _{b\left( {\gamma_n} \right)   } \right\}  \\
&\propto \mathcal{N} (  {x_n}  |{ 0 },\hat{\gamma}_n^{-1} ).
\end{aligned}
\label{M_fx_x}
\end{equation}

\textbf{Remarks}: In the above derivation, we assume that all variables are real-valued. The results can be easily extended to the case of proper complex variables. In this case, the UTAMP-SBL algorithm is still the same except that the coefficient 2 in Line 12 is removed and the superscript "$^H$" represents the conjugate transpose operation.  

\section{Numerical results}

We compare UTAMP-SBL with the state-of-the-art AMP based SBL algorithm GGAMP-SBL algorithm  \cite{al2018gamp} with estimated noise variance and 3 times of the true noise variance (as suggested in  \cite{al2018gamp}). As a performance benchmark, the support-oracle MMSE bound \cite{al2018gamp} is also  included. We set $N=1000$ and $M=800$. The vector $\mathbf{x}$ is drawn from a Bernoulli-Gaussian distribution with a non-zero probability $\rho$. The signal to noise power ratio $\text{SNR} \triangleq 10 \log10  ( \mathrm{E}\left\| \mathbf{ \mathbf{A x}} \right\| ^2/\mathrm{E}\left\| \mathbf{ \mathbf{w}} \right\| ^2 )$ (dB). 
We use the normalized mean squared error $\text{NMSE} \triangleq 10\log10  (\frac{1}{T}\sum_t \left\| \hat{ \mathbf{x}}^{t}- { \mathbf{ x}^{t}}\right\| ^2/\left\| { \mathbf{x}^{t}}\right\| ^2 )(\text{dB})$ to evaluate the recovery performance, where $\hat{ \mathbf{ x}}^t$ is an estimate of ${ \mathbf{ x}}^t$, and $T$ is the number of Monte Carlo simulations. 
\begin{figure}[htbp]
	\centering
	\includegraphics[width=1\linewidth]{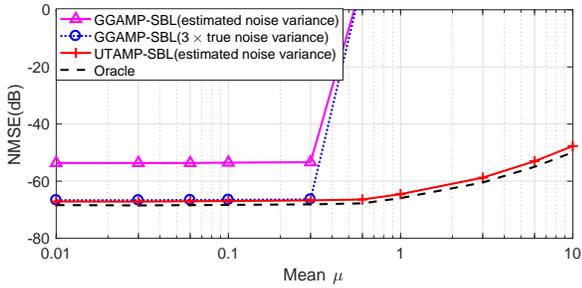}
	\caption{Performance under non-zero mean matrices ($\rho=0.1$).}
	\label{nonzeroresult}
\end{figure}
\begin{figure}[htbp]
	\centering
	\includegraphics[width=1\linewidth]{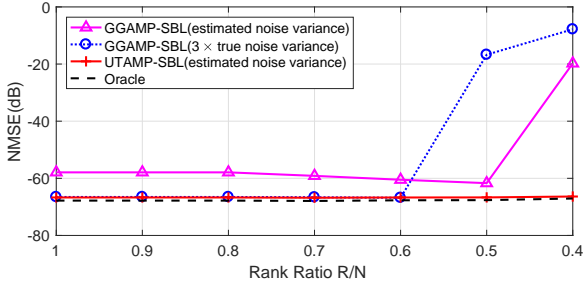}
	\caption{Performance under low rank matrices ($\rho=0.1$).}
	\label{lowrankresult}
\end{figure}

For UTAMP-SBL we set the maximum iteration number $t_{max}=300$ (noting that there is no inner iteration in UTAMP-SBL). For GGAMP-SBL, the maximum numbers of E-step and outer iteration are set to be 50 and 1000 respectively. The damping factor is 0.2. The SNR in all simulations is 60dB. All results are obtained by an ordinary PC with an Intel Core i7 3.50 GHz CPU and 64.0 GB RAM. We examine the performance of the algorithms with the following different types of measurement matrices. 
\subsubsection{Ill-Conditioned Matrix}
Matrix $\mathbf{ A }$ is constructed based on the SVD $\mathbf{A}=\mathbf{U \Lambda V}$ where $\bm{\Lambda}$ is a singular value matrix with $\Lambda_{i,i}/ \Lambda_{i+1,i+1} = \kappa^{1/(M-1)}$ (i.e., the condition number of the matrix is $\kappa$). The NMSE performance of all algorithms is shown in Fig.~\ref{conditionresult}.

\subsubsection{Correlated Matrix}
The correlated matrix $\textbf{A}$ is constructed using  $\textbf{A}=\textbf{C}_L^{1/2}\textbf{G}\textbf{C}_R^{1/2}$, where $\textbf{G}$ is an i. i. d. Gaussian matrix, and $\textbf{C}_L$ is an $M\times M$ matrix with the $(m,n)$th element given by $c^{|m-n|}$ where $c\in[0,1]$. Matrix $\textbf{C}_R$ is generated in the same way but with a size of $N\times N$. The parameter $c$ controls the correlation of matrix $\textbf{A}$. The NMSE performance of the algorithms with the parameter $c$ is shown in Fig.~\ref{correlationresult}.

\subsubsection{Non-Zero Mean Matrix} 
The elements of matrix $\mathbf{A}$ are drawn from a non-zero mean Gaussian distribution, i.e., $a_{m,n} \sim  \mathcal{N}( a_{m,n} | \mu, 1/N)$. The mean $\mu$ measures the derivation from the i. i. d. zero-mean Gaussian matrix. The NMSE performance of all the algorithms is presented in Fig.~\ref{nonzeroresult}.

\subsubsection{Low Rank Matrix} 
The measurement matrix $\textbf{A} =\textbf{BC}$, where the size of $\textbf{B}$ and $\textbf{C}$ are $M\times R$ and $R\times N$, respectively, and $R<M$. Both $\textbf{B}$ and $\textbf{C}$
are i.i.d. Gaussian matrices with mean zero and unit variance. The rank ratio $R/N$ is used to measure the deviation of matrix $\textbf{A}$ from the i.i.d. Gaussian matrix. The NMSE performance is shown in Fig.~\ref{lowrankresult}.

\begin{figure}[htbp]
	\centering
	\includegraphics[width=1\linewidth]{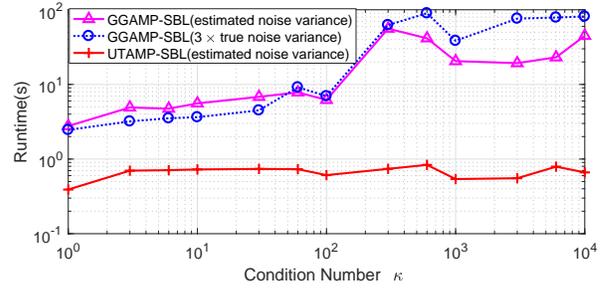}
	\caption{Runtime under ill-conditioned matrices ($\rho=0.3$).}
	\label{runtimerankresult}
\end{figure}
It can be seen from Fig.~\ref{conditionresult} to Fig.~\ref{lowrankresult}   that, when the deviation of the measurement matrices from the i.i.d. zero-mean Gaussian matrix is small, GGAMP-SBL (with $3\times$ true noise variance) and UTAMP-SBL deliver similar  performance, and both of them can approach the bound closely. However, when the deviation is relatively large, UTAMP-SBL can significantly outperform GGAMP-SBL, and in many cases, the performance of UTAMP-SBL is still close to the bound, which demonstrates that UTAMP-SBL is much more robust.    

Due to limited space, we only show the average running time for different algorithms for the case of ill-conditioned measurement matrices in Fig.~\ref{runtimerankresult}. It can be seen that UTAMP-SBL is much faster than GGAMP-SBL. 

\section{Conclusion}

In this work, we have proposed a UTAMP based SBL algorithm UTAMP-SBL, which inherits the low complexity and robustness of UTAMP to difficult measurement matrix $\mathbf{ A }$. It has been demonstrated that UTAMP-SBL can significantly outperform the state-of-the-art AMP based SBL algorithm in terms of robustness, speed and recovery accuracy for difficult measurement matrices. 


\bibliographystyle{IEEEtran}
\bibliography{IEEEabrv,mybib}

\begin{thebibliography}{10}
\providecommand{\url}[1]{#1}
\csname url@samestyle\endcsname
\providecommand{\newblock}{\relax}
\providecommand{\bibinfo}[2]{#2}
\providecommand{\BIBentrySTDinterwordspacing}{\spaceskip=0pt\relax}
\providecommand{\BIBentryALTinterwordstretchfactor}{4}
\providecommand{\BIBentryALTinterwordspacing}{\spaceskip=\fontdimen2\font plus
\BIBentryALTinterwordstretchfactor\fontdimen3\font minus
  \fontdimen4\font\relax}
\providecommand{\BIBforeignlanguage}[2]{{%
\expandafter\ifx\csname l@#1\endcsname\relax
\typeout{** WARNING: IEEEtran.bst: No hyphenation pattern has been}%
\typeout{** loaded for the language `#1'. Using the pattern for}%
\typeout{** the default language instead.}%
\else
\language=\csname l@#1\endcsname
\fi
#2}}
\providecommand{\BIBdecl}{\relax}
\BIBdecl

\bibitem{donoho2009message}
D.~L. Donoho, A.~Maleki, and A.~Montanari, ``Message-passing algorithms for
  compressed sensing,'' \emph{Proceedings of the National Academy of Sciences},
  vol. 106, no.~45, pp. 18\,914--18\,919, 2009.

\bibitem{javanmard2013state}
A.~Javanmard and A.~Montanari, ``State evolution for general approximate
  message passing algorithms, with applications to spatial coupling,''
  \emph{Information and Inference: A Journal of the IMA}, vol.~2, no.~2, pp.
  115--144, 2013.

\bibitem{rangana2019convergence}
S.~Rangana, P.~Schniterb, A.~K. Fletcherc, and S.~Sarkar, ``On the convergence
  of approximate message passing with arbitrary matrices,'' \emph{IEEE
  Transactions on Information Theory}, 2019.

\bibitem{manoel2014sparse}
A.~Manoel, F.~Krzakala, E.~W. Tramel, and L.~Zdeborov{\'a}, ``Sparse estimation
  with the swept approximated message-passing algorithm,'' \emph{arXiv preprint
  arXiv:1406.4311}, 2014.

\bibitem{vila2015adaptive}
J.~Vila, P.~Schniter, S.~Rangan, F.~Krzakala, and L.~Zdeborov{\'a}, ``Adaptive
  damping and mean removal for the generalized approximate message passing
  algorithm,'' in \emph{2015 IEEE International Conference on Acoustics, Speech
  and Signal Processing (ICASSP)}.\hskip 1em plus 0.5em minus 0.4em\relax IEEE,
  2015, pp. 2021--2025.

\bibitem{guo2015approximate}
Q.~Guo and J.~Xi, ``Approximate message passing with unitary transformation,''
  \emph{arXiv preprint arXiv:1504.04799}, 2015.

\bibitem{rangan2017vector}
S.~Rangan, P.~Schniter, and A.~K. Fletcher, ``Vector approximate message
  passing,'' in \emph{2017 IEEE International Symposium on Information Theory
  (ISIT)}.\hskip 1em plus 0.5em minus 0.4em\relax IEEE, 2017, pp. 1588--1592.

\bibitem{ma2017orthogonal}
J.~Ma and L.~Ping, ``Orthogonal amp,'' \emph{IEEE Access}, vol.~5, pp.
  2020--2033, 2017.

\bibitem{al2014sparse}
M.~Al-Shoukairi and B.~Rao, ``Sparse bayesian learning using approximate
  message passing,'' in \emph{2014 48th Asilomar Conference on Signals, Systems
  and Computers}.\hskip 1em plus 0.5em minus 0.4em\relax IEEE, 2014, pp.
  1957--1961.

\bibitem{al2018gamp}
M.~Al-Shoukairi, P.~Schniter, and B.~D. Rao, ``A gamp-based low complexity
  sparse bayesian learning algorithm,'' \emph{IEEE Transactions on Signal
  Processing}, vol.~66, no.~2, pp. 294--308, 2018.

\bibitem{guo2013iterative}
Q.~Guo, D.~D. Huang, S.~Nordholm, J.~Xi, and Y.~Yu, ``Iterative frequency
  domain equalization with generalized approximate message passing,''
  \emph{IEEE Signal Processing Letters}, vol.~20, no.~6, pp. 559--562, 2013.

\bibitem{fink1997compendium}
D.~Fink, ``A compendium of conjugate priors,'' \emph{See http://www. people.
  cornell. edu/pages/df36/CONJINTRnew\% 20TEX. pdf}, vol.~46, 1997.

\end{thebibliography}

\end{document}